\documentclass[rmp,aps,twocolumn]{revtex4}

\def\bio#1#2#3{{\it Biophys. J. }{\bf #1}, #2-#3}
\def\cell#1#2#3{{\it Cell } {\bf #1}, #2-#3}

\def\jcp#1#2#3{{\it J. Chem. Phys. }{\bf #1}, #2-#3}
\def\nat#1#2#3{{\it Nature }{\bf #1}, #2-#3}

\def\pnas#1#2#3{{\it Proc. Natl. Acad. Sci. USA }{\bf #1}, #2-#3}

\def\sci#1#2#3{{\it Science }{\bf #1}, #2-#3}

\usepackage{amsmath}
\usepackage{epsfig}
\begin{document}

\title{Force Modulating Dynamic Disorder: Physical Theory of
Catch-slip bond Transitions in Receptor-Ligand Forced Dissociation
Experiments}
\author{Fei Liu$^{1\ast}$ and Zhong-can Ou-Yang$^{1,2}$}
\address{$^{1}$Center for Advanced Study, Tsinghua University, Beijing, China}
\address{$^{2}$Institute of Theoretical Physics,
The Chinese Academy of Sciences, P.O.Box 2735 Beijing 100080,
China}
\email[Email address:]{liufei@tsinghua.edu.cn}
\date{\today}

\begin{abstract}
Recently experiments showed that some adhesive receptor-ligand
complexes increase their lifetimes when they are stretched by
mechanical force, while the force increase beyond some thresholds
their lifetimes decrease. Several specific chemical kinetic models
have been developed to explain the intriguing transitions from the
``catch-bonds" to the ``slip-bonds". In this work we suggest that
the counterintuitive forced dissociation of the complexes is a
typical rate process with dynamic disorder. An uniform
one-dimension force modulating Agmon-Hopfield model is used to
quantitatively describe the transitions observed in the single
bond P-selctin glycoprotein ligand 1(PSGL-1)$-$P-selectin forced
dissociation experiments, which were respectively carried out on
the constant force [Marshall, {\it et al.}, (2003) Nature {\bf
423}, 190-193] and the force steady- or jump-ramp [Evans {\it et
al.}, (2004) Proc. Natl. Acad. Sci. USA {\bf 98}, 11281-11286]
modes. Our calculation shows that the novel catch-slip bond
transition arises from a competition of the two components of
external applied force along the dissociation reaction coordinate
and the complex conformational coordinate: the former accelerates
the dissociation by lowering the height of the energy barrier
between the bound and free states (slip), while the later
stabilizes the complex by dragging the system to the higher
barrier height (catch).

\end{abstract}
\maketitle

Adhesive receptor-ligand complexes with unique kinetic and
mechanical properties paly key roles in cell aggregation, adhesion
and other life's functions in cells. A well studied example is the
receptors in selectin family which comprises E-, L- and P-selectin
interacting and forming ``bonds" with their glycoprotein ligands.
These bonds are primarily responsible for the tethering and
rolling of leukocytes on inflamed endothelium under shear
stress~\cite{McEver,Konstantopoulos98}. In particular, in the past
two years great experimental
efforts~\cite{Marshall,Evans2004,Sarangapani,Yago,Marshall05} have
been devoted to study the surprising kinetic and mechanical
behaviors of the bonds between L- and P-selectin and P-selectin
glycoprotein ligand 1 (PSGL-1) at the single molecule level: the
lifetimes of these bonds first increase with initial application
of small force, which are termed ``catch" bonds, and subsequently
decrease, which are termed ``slip" bonds when the force increases
beyond some thresholds. The most important biological meaning of
this discovery is that the catch-slip transitions of the
PSGL-1$-$L- and $-$P-selectin bonds may provide a direct
experimental evidence at the single-molecule level to account for
the shear threshold effect~\cite{Finger,Lawrence}, in which the
number of rolling leukocytes first increases and then decreases
while monotonically increasing shear stress.

On the theoretical side, it is a challenge to give a reasonable
physical theory or model to explain the counterintuitive bond
transitions. Bell~\cite{Bell} firstly suggested that the force
induced dissociation rate of adhesive receptor-ligand complex
could be described by,
\begin{eqnarray}
k_{\rm off}(f)=k^0_{\rm off}\exp[f\xi^{\ddag}/k_{\rm B}T],
\label{Bellmodel}
\end{eqnarray}
where $k^0_{\rm off}$ is the intrinsic dissociation rate constant
in the absence of force, $\xi^{\ddag}$ is the distance from the
bound state to the energy barrier, $f$ is a projection of external
applied force along the dissociation coordinate, $k_{\rm B}$ the
Boltzmann's constant, and $T$ is absolute temperature. The
validity of the model has been demonstrated in
experiments~\cite{Alon,Chen}. Although later at least four models
have been put forward to explain and understand various
receptor-ligand forced dissociation
experiments~\cite{Dembo,Dembo94,Evans97}, they cannot predict
catch bonds because force in these models only lowers height of
the energy barrier while shortening lifetimes of the bonds. An
exception is the Hookean spring model proposed by Dembo many years
ago~\cite{Dembo,Dembo94}, in which a catch bond was raised in
mathematics. Compared to the exponential decay of the lifetimes of
slip-bonds with respect to force in experiments~\cite{Chen}, the
model claimed that the lifetimes decrease exponentially with the
square of force~\cite{Dembo}. In addition, the Hookean spring
model cannot account for the catch-slip bond transitions in
self-consistent term. Prompted by the intriguing experimental
observations, three chemical kinetic models have been developed.
Evans {\it et al.} presented a two pathways, two bound states
model with rapid equilibrium assumption between the two states.
They suggested that the catch-slip bond transitions take place due
to applied force switching the pathways from the one with slower
dissociation rate to the fast one~\cite{Evans2004}. Although this
insightful viewpoint well described force jump-ramp experiments,
there are two apparent flaws in physics. First if the forced
dissociation experiments were performed at very low temperatures
or higher solvent viscosities, force would be independent of the
bond dissociations since the force in the two pathways and two
bound states model only acts on the inner bound states, while
these states would be ``frozen" under this circumstances. The
other is that force does not accelerate the dissociation processes
further when the force is sufficiently large for the fast
dissociation rate is a constant. The next model given by Barsegov
and Thirumalai~\cite{Barsegov} with same kinetic scheme seems to
improve the two flaws in which the dissociation rates of the two
pathways were allowed to be force-dependent with the Bell
formulas. Unfortunately, so many independent reaction constants
with arbitrary dependencies on the force parameters (total seven
parameters) and the final dissociation rate depending on them in a
complicated way make the physical explanations and the
determination of the parameters difficult to track. Very recently,
a competitive two pathways and one bound state model was proposed
by Thomas {\it et al.}~\cite{Pereverzev}. This model is distinct
from the others because there is a catch pathway therein, which
was thought to arise from a backward unbinding pathway. But the
model is not intuitively obvious just like the authors pointed
out.

As one type of noncovalent bonds, interactions of adhesive
receptors and their ligands are weaker. Moreover, the interfaces
between them have been reported to be broad and shallow, such as
the crystal structure of the PSGL-1$-$P-selectin bond
revealed~\cite{Somers}. Therefore it is plausible that the energy
barriers for the bonds are fluctuating with time due to either
global conformational changes or local conformational changes at
the interfaces. Association/dissociation reactions with
fluctuating energy barrier have been deeply studied by statistical
physicists in terms of rate processes with \emph{dynamic
disorder}~\cite{Zwanzig90} during the past two decades. A
prototype is the ligand rebinding in myoglobin where the rate
constant depends on a protein coordinate~\cite{Agmon}. Hence it is
of interest to determine whether the fluctuation of energy barrier
responses to catch-slip bond transitions. Such studies should be
meaningful since in the Bell's initial work and the other models
developed later, the intrinsic rate constants $k^0_{\rm off}$ were
deterministic and time-independent. It is possible to derive
unexpected results from the relaxation of this restriction.
Stimulated by the two considerations, in the present work we
propose that the intrinsic dissociation rate in the Bell model is
controlled by a conformational coordinate of receptor-ligand
complex, while the coordinate is fluctuating as a Brownian motion
in a bound harmonic potential; applied force not only lowers the
height of the energy barrier as described in Eq.~\ref{Bellmodel}
but also modulates the distribution of the conformational
coordinate. In addition to well predicting the experimental data,
our theory may also provide a new physical mechanism for the
dissociation rates suggested by Bell~\cite{Bell} and
Dembo~\cite{Dembo} early.

\section{Theory and methods}
The physical picture of our theory for the forced dissociation of
receptor-ligand bonds is very similar with the small ligand
binding to heme proteins~\cite{Agmon}: there is a energy surface
for dissociation which dependents on both the reaction coordinate
for the dissociation and the conformational coordinate $x$ of the
complex, while the later is perpendicular to the former; for each
conformation $x$ there is a different dissociation rate constant
which obeys the Bell rate model, while the distribution of $x$
could be modulated by the force component along x-direction;
higher temperature or larger diffusivity (low viscosities) allows
$x$ variation within the complex to take place, which results in a
variation of the energy barrier of the bond with time.

There are two types of experimental setups to measure forced
dissociation of receptor-ligand complexes. First we consider
constant force mode~\cite{Marshall,Sarangapani}. A diffusion
equation in the presence of a coordinate dependent reaction is
given by~\cite{Agmon}
\begin{eqnarray}
 \label{origindiffusionequation}
\frac{\partial p(x,t)}{\partial t}=D\frac{\partial^2 p}{\partial
x^2}+ D\beta \frac{\partial}{\partial x}\left(p\frac{\partial
V_{f_\perp}}{\partial x}\right)-k_{\rm off}(x,f_\parallel)p,
\end{eqnarray}
where $p(x,t)$ is probability density for finding a value $x$ at
time $t$, and $D$ is the diffusion constant. The motion is under
influence of a force modulating potential $V_{f_\perp}(x)=V_{\rm
i}(x)-f_\perp x$, where $V_{\rm i}(x)$ is intrinsic potential in
the absence of any force, and a coordinate-dependent Bell rate. In
the present work Eq.~\ref{Bellmodel} depends on $x$ through the
intrinsic rate $k^0_{\rm off}(x)$, and the distance $\xi^\ddag$ is
assumed to be a constant for simplicity. Here $f_\perp$ and
$f_\parallel$ are respective projections of external force $f$
along the reaction and conformational diffusion coordinates:
\begin{eqnarray}
f_\perp&=&f\sin\theta,\\
f_\parallel&=&f\cos\theta\ge 0,\nonumber
\end{eqnarray}
and $\theta$ is the angle between $f$ and the reaction coordinate.
We are not ready to study general potentials here. Instead, we
focus on specific $V_{\rm i}(x)$s, which make $V_{f_\perp}(x)$ to
be
\begin{eqnarray}
V_{f_\perp}(x)=V\left(x-\eta-\frac{f_{\perp}}{\kappa}\right)+W(f_{\perp}),
\label{forcedependentpotential}
\end{eqnarray}
where $\eta$ and $\kappa$ are two constants with Length and Force
dimensions. For example for a harmonic potential
\begin{eqnarray}
V_{\rm i}(x)=V_0+k_x(x-x_0)^2/2 \label{harmonicpotential}
\end{eqnarray}
with a spring constant $k_x$ in which we are interested, it gives
\begin{eqnarray}
V\left(x-x_0-\frac{f_{\perp}}{k_x}\right)=
\frac{k_x}{2}\left(x-x_0-\frac{f_{\perp}}{k_x}\right)^2
\end{eqnarray}
and
\begin{eqnarray}
W(f_\perp)=V_0-f_\perp x_0-\frac{f_\perp^2}{2k_x}.
\end{eqnarray}
Defining a new coordinate variable $y=x-\eta-f_\perp/\kappa\rm$,
we can rewrite Eq.~\ref{origindiffusionequation} with the specific
potentials into
\begin{eqnarray}
\frac{\partial \rho(y,t)}{\partial t}&=&D\frac{\partial^2
\rho}{\partial y^2}+  D\beta \frac{\partial}{\partial
y}\left(\rho\frac{\partial V(y)}{\partial y}\right)-k_f(y)\rho
\label{newdiffusionequation}
\end{eqnarray}
where $k_f(y)=k_{\rm off}(y+\eta+f_\perp/\kappa,f_\parallel)$.
Compared to the original work by Agmon and Hopfield~\cite{Agmon},
our problem for the constant force case is almost same except the
reaction rate now is a function of the force. Hence, all results
obtained previously could be inherited with minor modifications.
Considering the requirement of extension of
Eq.~\ref{origindiffusionequation} to dynamic force in the
following, we present the essential definitions and calculations.

Substituting
\begin{eqnarray}
\rho(y,t)=N_0\exp\left(-\frac{V}{k_{\rm B}T}\right)\phi(y,t)
\label{transform}
\end{eqnarray}
into Eq.~\ref{newdiffusionequation}, one can convert the
diffusion-reaction equation into Schr$\ddot o$dinger-like
presentation~\cite{Kampen}.
\begin{eqnarray}
\frac{\partial \phi}{\partial t}=D\frac{\partial^2 \phi}{\partial
y^2}-U_f(y)\phi=-{\cal H}_f(\phi),\label{Schodingerequation}
\end{eqnarray}
where $N_0$ is the normalization constant of the density function
at $t=0$, and the ``effective" potential
\begin{eqnarray}
\label{forcedependentquantumpotential}
U_f(y)&=&U(y)+k_f(y)\\
&=&\frac{D}{2k_{\rm B}T}\left[\frac{1}{2k_{\rm
B}T}\left(\frac{\partial V}{\partial y}\right)^2-\frac{\partial^2
V}{\partial y^2}\right]+k_f(y).\nonumber
\end{eqnarray}
We define $U(y)$ for it is independent of the force $f$.
Eq.~\ref{Schodingerequation} can be solved by eigenvalue
technique~\cite{Agmon}. At larger $D$ in which we are interested
here, only the smallest eigenvalue $\lambda_0(f)$ mainly
contributes to the eigenvalue expansion which is obtained by
perturbation approach~\cite{Messiah}: if the eigenfunctions and
eigenvalues of the ``unperturbed" Schr$\ddot o$dinger operator
\begin{eqnarray}
{\cal H}=-\frac{\partial^2}{\partial y^2}+U(y)
\end{eqnarray}
in the absence of $k_f(y)$ have been known,
\begin{eqnarray}
{\cal H}\phi^0_n=-\lambda^0_n\phi^0_n,
\end{eqnarray}
and $k_f$ is adequately small, the first eigenfunction $\phi_0(f)$
and eigenvalue $\lambda_0(f)$ of the operator ${\cal H}_f$ then
are respectively given by
\begin{eqnarray}
\label{eigenfunctionexpansion}
\phi_0(f)&=&\phi^{(0)}_0+\phi^{(1)}_0(f)+\cdots\\
&=&\phi^0_0+\sum_{m\neq
0}\frac{\int\phi^0_0(y)k_f(y)\phi^0_m(y)dy}{\lambda^0_0-\lambda^0_m}\phi^0_m+\cdots\nonumber
\end{eqnarray}
and
\begin{eqnarray}
\label{eigenvalueexpansion}
\lambda_0(f)&=&\lambda^{(0)}_0+\lambda^{(1)}_0(f)+\lambda^{(2)}_0(f)+\cdots\\
&=&\lambda^0_0+\int\phi^0_0(y)k_f(y)\phi^0_0(y)dy+\nonumber\\
&&\sum_{m\neq 0}\frac{\left(\int \phi^0_0(y)k_f(y)
\phi^m_0(y)dy\right)^2}{\lambda^0_0-\lambda^0_m}+\cdots.\nonumber
\end{eqnarray}
Considering that the system is in equilibrium at the initial time,
{\it i.e.}, no reactions at the beginning, the first eigenvalue
$\lambda^{(0)}_0$ must vanish. On the other hand, because
\begin{eqnarray}
\phi_0^0(y)\propto \exp\left(-V(y)/2k_{\rm B}T\right),
\end{eqnarray}
and the square of $\phi_0^0$ is just the equilibrium Boltzmann
distribution $p_{\rm eq}(y)$ with the potential $V(y)$, we
rewritten the first correction of $\lambda_0(f)$ as
\begin{eqnarray}
\label{equilibriumdistribution}
\lambda^{(1)}_0(f)=\int p_{\rm eq}(y)k_f(y)d    y,\\
p_{\rm eq}(y)\propto \exp\left[-V(y)/2k_{\rm B}T\right].\nonumber
\end{eqnarray}
Substituting the above formulaes into Eq.~\ref{transform}, the
probability density function then is approximated to
\begin{eqnarray} \rho(y,t)\approx
N_0\exp\left(-\frac{V}{2k_{\rm
B}T}\right)\exp[-\lambda_0(f)t]\phi_0(f)
\end{eqnarray}

The quantity measured in the constant force experiments is the
mean lifetime of the bond $\langle\tau\rangle$,
\begin{eqnarray}
\label{averagelifetime}
\langle\tau\rangle=-\int_0^{\infty}t\frac{dQ}{dt}dt=\int_0^{\infty}Q(t)dt,
\end{eqnarray}
where the survival probability $Q(t)$ related to the probability
density function is given by
\begin{eqnarray}
\label{survivalprobability}
Q(t)&=& \int  p(x,t)dx=\int \rho(y,t)dy\nonumber\\
&\approx&\exp\left[-t\left(\lambda_0^{(1)}(f)+\lambda_0^{(2)}(f)\right)\right].
\end{eqnarray}

In addition to the constant force mode, force could be
time-dependent, {\it e.g.}, force increasing with a constant
loading rate in biomembrane force probe (BFP)
experiment~\cite{Evans2004}. In principle the scenario would be
more complicated than that for the constant force mode. We assume
that the force is loaded slowly compared to diffusion-reaction
process. We then make use an adiabatic approximation analogous to
what is done in quantum mechanics. The correction of this
assumption would be tested by the agreement between theoretical
calculation and experimental data. We still use
Eq.~\ref{origindiffusionequation} to describe bond dissociations
with the dynamic force, therefore we obtain the almost same
Eqs.~\ref{forcedependentpotential}-\ref{forcedependentquantumpotential}
except that the force therein is replaced by a time-dependent
function $f_t$. We immediately have~\cite{Messiah}
\begin{eqnarray}
\phi(y,t)\approx \exp\left[-\int_0^t \left(\lambda_0(f_{t'})+
B(t')\right)dt' \right]\phi_0(f_t),
\end{eqnarray}
where the ``Berry phase"
\begin{eqnarray}
B(t)=\int\phi_0(f_t) \frac{\partial}{\partial t}\phi_0(f_t)dy,
\end{eqnarray}
and $\phi_0(f_t)$ is the first eigenfunction of the time-dependent
Sch$\ddot o$dinger operator
\begin{eqnarray}
{\cal H}_{f_t}&=&{\cal H} + k_{f_t}(y).
\end{eqnarray}
Because the eigenvalues and eigenfunctions of the above operator
cannot be solved analytically for general $k_{f_t}$, we also apply
the perturbation approach. Hence, we obtain $\phi_0(f_t)$ and
$\lambda_0(f_t)$ by replacing $k_f$ in
Eqs.~\ref{eigenfunctionexpansion} and \ref{eigenvalueexpansion}
with $k_{f_t}$. The Berry phase then is approximated to
\begin{eqnarray}
\label{Berryphase}
B(f_t)&\approx&\sum_{m\neq0}\left(\frac{1}{\lambda^0_m}\right)^2\int\phi^0_0(y)k_{f_t}(y)\phi^0_m(y)
dy\times\nonumber\\
&&\int \phi^0_0(y)\frac{dk_{f_t}}{dt}\phi^0_m(y) dy
\end{eqnarray}
Finally, the survival probability for the dynamic force is given
by
\begin{eqnarray}
&&Q(t)\approx\exp\left[-\int_0^t\left(\lambda^{(1)}_0(f_{t'})+
\lambda^{(2)}_0(f_{t'})+B(f_{t'})\right)dt'\right]\nonumber\\
\end{eqnarray}

Different from the constant force mode, data of the dynamic force
experiments is typically presented in terms of the force
histogram, which corresponds to the probability density of the
dissociation forces $p(f)$
\begin{eqnarray}
p(f)=-\frac{dQ}{dt}\left/\frac{df}{dt}\right.
\end{eqnarray}
Particularly, when the force is a linear function of time
$f=f_0+rt$, where $r$ is the loading rate, and zero or nonzero of
$f_0$ respectively corresponds to the steady- or jump-ramp force
mode in the dynamic force experiment~\cite{Evans2004}, we have
\begin{eqnarray}
\label{ruputureforcedistribution} &&P(f,f_0)\approx\frac{1}{r}
\left[\lambda^{(1)}_0(f)+
\lambda^{(2)}_0(f)+B(f)\right]\times\\
&&\exp\left[-\frac{1}{r}\int_{f_0}^f\left(\lambda^{(1)}_0(f')+
\lambda^{(2)}_0(f')+ B(f')\right)df'\right].\nonumber
\end{eqnarray}

\section{Results}
We consider a bounded diffusion in the harmonic potential
Eq.~\ref{harmonicpotential}. Then $\cal H$ reduces to a harmonic
oscillator operator with
\begin{eqnarray}
U(y)=\frac{Dk_x}{2k_{\rm B}T}\left(\frac{k_xy^2}{2k_{\rm
B}T}-1\right).
\end{eqnarray}
Its eigenvalues and eigenfunctions are
\begin{eqnarray}
\label{unpertubatedeigenvaluesharmonic} \lambda^0_n&=&nDk_x/k_{\rm
B}T
\end{eqnarray}
and
\begin{eqnarray}
\phi^0_n(z)&=&2^{-n/2}\pi^{-1/4}(n!)^{-1/2}e^{-z^2/2}H_n(z),
\end{eqnarray}
respectively, where $z=(k_x/2k_{\rm B}T)^{1/2}y$ and $H_n(z)$ is
the Hermite polynormials~\cite{Messiah}. Given that the intrinsic
dissociation rate satisfies the Arrenhenius form
\begin{eqnarray}
k_{\rm off}^0(x)=k_0\exp\left[-\Delta G^\ddag(x)/k_{\rm
B}T\right],
\end{eqnarray}
where the height of the energy barrier along the reaction
coordinate $\Delta G^\ddag(x)$ is a function of the conformational
coordinate $x$. According to the form of barrier, we first analyze
two simple and meaningful cases.
\\
{\bf Bell-like forced dissociations}. The simplest function of the
energy barrier might be linear with respect to $x$,
\begin{eqnarray}
\label{linearheightfunction} \Delta G^\ddag(x)=\Delta
G^\ddag_0+k_g(x-x_0),
\end{eqnarray}
where $\Delta G^\ddag_0$ is the height at position $x_0$, and the
slope $k_g\ge0$ (its dimension Force) for the perturbation
requirement in solving Eq.~\ref{Schodingerequation}. According to
Eqs.~\ref{eigenvalueexpansion} and \ref{Berryphase}, we easily get
\begin{eqnarray}
\label{secondcorrectionsforlinearpotential}
\lambda^{(1)}_0(f)&=&k_0\exp\left[-\beta \Delta
G^\ddag_0+\frac{\beta k_g^2}{2k_x}\right]\times\nonumber\\
&&\exp\left[\beta\left(\xi^\ddag f_\parallel-\frac{k_g}{k_x}f_\perp\right)\right],\nonumber\\
\lambda^{(2)}_0(f)&=&\frac{-k_0^2}{\beta
Dk_x}\exp{\left[-2\beta\Delta G^\ddag_0+\frac{\beta
k_g^2}{k_x}\right]}\times\\
&&\exp{\left[2\beta\left(\xi^\ddag f_\parallel-\frac{
k_g}{k_x}f_\perp\right)\right]}\sum_{n=1}^{\infty}\frac{1}{nn!}\left(\frac{\beta
k_g^2}{k_x}\right)^n,\nonumber
\end{eqnarray}
and
\begin{eqnarray}
\label{Berryexpressionforlinearpotential}
B(f_t)&=&\frac{d}{dt}\left(\xi^\ddag
{f_t}_\parallel-\frac{k_g}{k_x}{f_t}_\perp\right)\times\nonumber\\
&&\frac{k_0^2}{\beta D^2k_x^2} \exp{\left[-2\beta\Delta
G^\ddag_0+\frac{\beta
k_g^2}{k_x}\right]}\times\\
&&\exp{\left[2\beta\left(\xi^\ddag {f_t}_\parallel-\frac{
k_g}{k_x}{f_t}_\perp\right)\right]}
\sum_{n=1}^{\infty}\frac{1}{n^2n!}\left(\frac{\beta
k_g^2}{k_x}\right)^n,\nonumber
\end{eqnarray}
where $\beta=1/k_{\rm B}T$. For large $D$ or $k_x$ (or very small
$T$), the second correctness and the Berry phase tend to zero.
Under these limitations the first eigenvalue of
Eq.~\ref{origindiffusionequation} is approximated to be
\begin{eqnarray}
\label{correctedBellform} \lambda_0(f)\approx k_0\exp\left[-\beta
\Delta G^\ddag_0+\frac{\beta k_g^2}{2k_x}\right] \exp[\beta
d^\ddag f].
\end{eqnarray}
Here we define a new distance
\begin{eqnarray}
\label{newdistance} d^\ddag=\xi^\ddag \cos\theta-\zeta \sin\theta,
\end{eqnarray}
where $\zeta=k_g/k_x$ whose dimension is Distance. We see that the
presence of the complex conformational coordinate could modify the
original Bell model in novel ways: (i) $d^\ddag>0$,
Eq.~\ref{correctedBellform} is indistinguishable from the origin
Bell model, although the projection distance $\xi^\ddag\cos\theta$
from the bound state to the energy barrier may be increased or
decreased in terms of the orientation of the applied force. In
particular, if the force is antiparallel to $x$, {\it i.e.},
$\sin\theta=-1$, we get a Bell-like rate expression with a
``distance" $\zeta$; (ii) $d^\ddag=0$, the force does not affect
dissociations of the bonds, which have been named ``ideal" bonds
by Dembo~\cite{Dembo,Dembo94}; (iii) $d^\ddag <0$, the force slows
down dissociations of the bonds. It is ``catch" bonds in which we
are interested. In contrast to the catch behavior suggested by
Dembo~\cite{Dembo}, the rate decays exponentially with respect to
the force instead of the square of the force. Given the linear
function Eq.~\ref{linearheightfunction} and $f_{\perp}>0$,
increasing of the force only stabilizes the bonds by dragging the
system to the higher energy barriers (catch), whereas the other
force component $f_{\parallel}$ destabilizes the complex by
lowering the energy barriers (slip). Therefore the sign of the
distance $d^\ddag$ in fact reflects a competition of the
two contrast effects of the same force. \\
{\bf Dembo-like forced dissociations}. Another function of the
energy barrier is a harmonic with a spring constant $k_g$
\begin{eqnarray}
\Delta G^\ddag(x)=\Delta G^\ddag_1+k_g(x-x_1)^2/2,
\end{eqnarray}
where $\Delta G^\ddag_1$ is the barrier height at position $x_1$.
Because for any form of the barrier height, the dependence of
$\lambda_0^{(2)}$ and $B(f_t)$ on $D$ is the same from
Eq.~\ref{unpertubatedeigenvaluesharmonic}, we only consider the
large $D$ limitation in the following. Hence we have
\begin{eqnarray}
\lambda_0(f)\approx
k_0\left(\frac{k_x}{k_x+k_g}\right)^\frac{1}{2}\exp\left[-\beta\Delta
G^\ddag_1+\beta \xi^\ddag f_\parallel\right]\times\nonumber\\
\exp\left[-\frac{\beta
k_g(f_\perp-k_x(x_1-x_0))^2}{2k_x(k_x+k_g)}\right]
\end{eqnarray}
Given $\sin\theta>0$ and $x_1-x_0>0$, we find that there is a
interesting transition from slip to catch bond when the force
increase over a threshold $k_x(x_1-x_0)/\sin\theta$; otherwise
only catch bond presents. We note that the latter is very similar
to the result proposed by Dembo~\cite{Dembo} even their physical
origins are completely different: both of them exponentially
dependent on the square of the force.\\
{\bf Comparison with the experiments.}
\begin{figure}[htpb]
\begin{center}
\includegraphics[width=0.9\columnwidth]{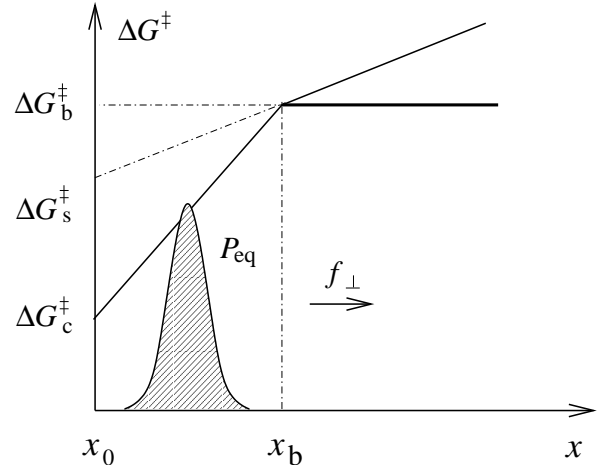}
\caption{Schematic diagram of the height function of the energy
barrier with respect to the coordinate $x$. $\Delta G^\ddag_s$ and
$\Delta G^\ddag_c$ are the value of the linear functions in
Eq.~\ref{bendfunction} at position $x_0$, while $\Delta G^\ddag_b$
is the intersection of the functions at $x_b$. The bold solid line
is the minimum model which is used to fit the experiment. The
shaded area represents the equilibrium distribution of the
conformational coordinate under the potential $V(y)$ (
Eq.~\ref{equilibriumdistribution}).} \label{figure1}
\end{center}
\end{figure}
In the constant force rupture experiment of the
PSGL-1$-$P-selectin complex, the dissociation rate as the inverse
mean lifetime of the complex first decreased and then increased
when the applied force increased beyond a force
threshold~\cite{Marshall}. We now can easily understand this
counterintuitive transition according to the previous discussion
about the Bell-like dissociation rate: the dissociation effect of
$f_{\parallel}$ regains its dominance when the force is beyond the
threshold. Although in principle we can construct various barrier
height functions which result into catch-slip transitions, the
most simplest form may be a composition of two linear functions
\begin{eqnarray}
\label{bendfunction} \Delta G^\ddag(x)=\left\{
\begin{gathered}
 \Delta G^{\ddag}_c(x)=\Delta G^{\ddag}_b+k_c(x-x_b),  {\text { }x\le x_b}\hfill \\
 \Delta G^{\ddag}_s(x)=\Delta G^{\ddag}_b+k_s(x-x_b),  {\text { }x>x_b}\hfill \\
\end{gathered}  \right.
\end{eqnarray}
where we require that the distances defined in
Eq.~\ref{newdistance} with $k_c$ and $k_s$ are respectively minus
and positive. For convenience, their absolute values are
correspondingly denoted by $d^\ddag_c$ and $d^\ddag_s$. Define two
``intrinsic" dissociation constants
\begin{eqnarray}
k^c_0=k_0\exp[-\beta\Delta G^\ddag_c(x_0)],\nonumber\\
k^s_0=k_0\exp[-\beta\Delta G^\ddag_s(x_0)].
\end{eqnarray}
Fig.~\ref{figure1} shows the characteristics of the function. We
then have
\begin{eqnarray}
\label{reactionratebendlandscape}
\lambda_0(f)&\approx&\frac{k_0^c}{2}\exp\left[\frac{\beta
k_c^2}{2k_x} \right]\exp\left[-\beta d^\ddag_c
f\right]\nonumber\\
&&\times{\rm
erfc}\left[-\left(\Delta+\frac{k_c}{k_x}\right)\sqrt{\frac{\beta
k_x}{2}}+f\sqrt{\frac{\beta}{2k_x}}\sin\theta \right] \nonumber\\
&&+\text{ }\frac{k_0^s}{2}\exp\left[\frac{\beta
k_s^2}{2k_x}\right]\exp\left[\beta d^\ddag_s
f\right]\\
&&\times{\rm
erfc}\left[\left(\Delta+\frac{k_s}{k_x}\right)\sqrt{\frac{\beta
k_x}{2}}-f\sqrt{\frac{\beta}{2k_x}}\sin\theta \right]\nonumber
\end{eqnarray}
where $\Delta =x_b-x_0$, and the complementary error function
\begin{eqnarray}
{\rm erfc}(x)=\frac{2}{\sqrt{\pi}}\int_x^\infty e^{-x^2}dx.
\end{eqnarray}
Before fixing numerical values of the parameters in
Eq.~\ref{reactionratebendlandscape}, we first simply analyze the
main properties of $\lambda_0(f)$ given $\Delta \gg 0$: (i) in the
absence of force, due to ${\rm erfc}(-\infty)=2$ and ${\rm
erfc}(+\infty)=0$, we have
\begin{eqnarray}
\label{smallforcelimit} \lambda_0\approx k_0^c\exp(\beta
k_c^2/2k_x),
\end{eqnarray}
which is the same with that obtained by Agmon and
Hopfield~\cite{Agmon}; (ii) if force is nonzero and smaller,
\begin{eqnarray}
\label{largeforcelimit} \lambda_0\approx k_0^c\exp(\beta
k_c^2/2k_x)\exp(-\beta d^\ddag_cf),
\end{eqnarray}
which means that the bond is catch; and finally (iii), when the
force is sufficiently large, Eq.~\ref{reactionratebendlandscape}
reduces to
\begin{eqnarray}
\lambda_0\approx k_0^s\exp(\beta k_s^2/2k_x)\exp(\beta
d^\ddag_sf).
\end{eqnarray}
It is the ordinary slip bond.

\begin{figure}[htpb]
\begin{center}
\includegraphics[width=0.9\columnwidth]{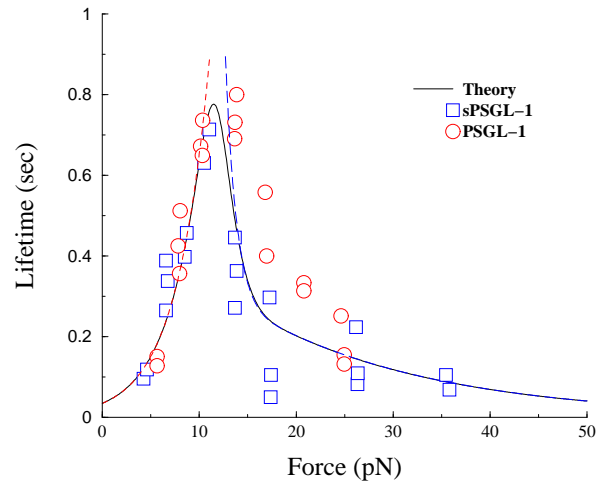}
\caption{The mean lifetime as a function of force for the bonds of
dimeric P-selectin with monomeric sPSGL-1 (square
symbols)~\cite{Marshall} and the rescaled dimeric PSGL-1 (circle
symbols) from Ref.~\cite{Pereverzev}. The two dash curves are
respectively calculated by the two addition terms in
Eq.~\ref{reactionratebendlandscape} with the same parameters.}
\label{figure2}
\end{center}
\end{figure}

There are total eight independent parameters presenting in
Eq.~\ref{reactionratebendlandscape}: $\theta,\text{ }k_x,\text{
}k_c,\text{ }k_s,\text{ }\Delta,\text{ }\xi^\ddag,\text{ }k_0$,
and $\Delta G^\ddag_c(x_0)$. It is not necessary to determine all
of them, which is also impossible only through fitting to the
experimental data~\cite{Marshall}. For example, the latter two
parameters are lumped into $k^c_0$, while and $\xi^\ddag$ always
presents with $\cos\theta$ together. What we really concern with
is the coefficients of the force and the factors before the error
functions in Eq.~\ref{reactionratebendlandscape}. They can be
obtained by least square fit. Guided by the properties
Eqs~\ref{smallforcelimit} and \ref{largeforcelimit}, the fitting
process in fact is simple. Even so, we are still able to fix all
parameters from the fitting results if we study a minimum model in
which the slop $k_s$ is zero and $\theta=\pi/6$. Here the
particular value of the angle is actually of no particular
significance and it is only as a reference. We immediately have:
$k_c\approx0.60\text{ pN}$, $k_x\approx0.21\text{ pN nm}^{-1}$,
$\xi^\ddag\approx0.25\text{ nm}$, $\Delta\approx33\text{ nm}$; the
other interesting parameters see Tab. 1, where the values are
independent of the angle.
\begin{table}
\caption{Comparison of the parameters of the present theory, and
the two-pathway and one energy-well model presented by Thomas et
al.~\cite{Pereverzev} on the constant force (cf) and the dynamic
force (df) modes. The parameters for the slip behavior of the
P-selectin are also listed as a reference~\cite{Hanley}. }
\begin{center}
\begin{tabular}{cccccc}
\hline &$d_c$ nm & $d_s$ nm &  $k_0^c$ sec$^{-1}$ & $k_0^s$ sec$^{-1}$&\\
\hline Experiment &&0.14&&0.2&\\
Dynamic disorder by us &1.2 & 0.22 &23.2 & 1.68 &\\
Two-pathway one-well (cf) & 2.2 & 0.5 & 120 & 0.25 & \\
Two-pathway one-well (df) & 0.4 & 0.2 & 20 & 0.34 & \\
\hline
\end{tabular}
\label{table}
\end{center}
\end{table}
Substituting these values into Eq.~\ref{reactionratebendlandscape}
and according to Eq.~\ref{averagelifetime}, we calculate the mean
lifetime of the PSGL-1$-$P-selectin complex with respect to
different constant force in Fig.~\ref{figure2}: the agreement
between theory and the experimental data is quite good.

\begin{figure}[htpb]
\begin{center}
\includegraphics[width=0.9\columnwidth]{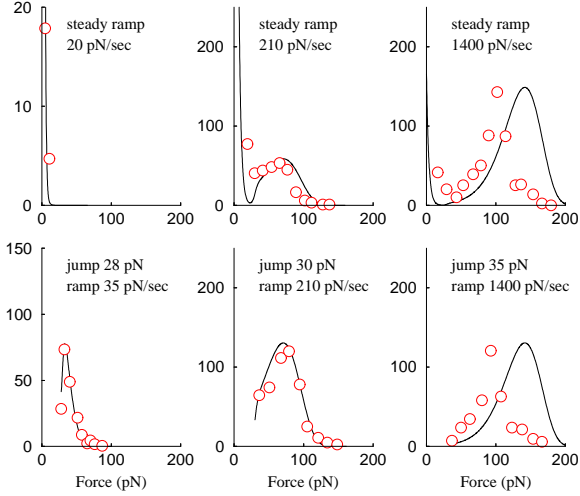}
\caption{The probability density of the dissociation forces
$P_{\rm d}(f,f_0)$ under the different loading rates predicted by
our theory (solid curves) for the PSGL-1$-$P-selectin complex. The
symbols are from the force steady- and jump-ramp experimental
data~\cite{Evans2004}. The apparent deviations between the theory
and the data in the last column may be from the invalidation of
the assumption of two equivalent bonds at higher loading rates.
}\label{figure3}
\end{center}
\end{figure}

More challenging experiments to our theory are the force steady-
and jump-ramp modes~\cite{Evans2004}. Under large $D$ limitation,
Eq.~\ref{ruputureforcedistribution} reduces to
\begin{eqnarray}
P(f,f_0)\approx\frac{\lambda_0(f)}{r}\exp \left
[-\frac{1}{r}\int_{f_0}^f\lambda_0(f')df' \right].
\label{approxruputureforcedistribution}
\end{eqnarray}
We see that the mean lifetime can be extracted from the above
equation by setting $f=f_0$, {\it i.e.}, $\langle \tau\rangle
=1/rP(f_0,f_0)$. We calculate the dissociation force distributions
of the steady- and jump-ramp modes at three loading rates to
compare with the BFP experiments performed by Evans {\it et
al.}~\cite{Evans2004}. Here we are not ready to fit the
experiments afresh; instead we directly apply the parameters
obtained from the constant force mode to current case. Because the
BFP experimental data is for {\it dimeric} ligand PSGL-1, whereas
our parameters are from monomeric ligand sPSGL-1. Therefore it is
necessary to map our predictions for the single bond
sPSGL-1$-$P-selectin to the double bonds. A natural assumption is
that the two bonds share the same force and fail randomly. The
same assumption has been used in previous
works~\cite{Pereverzev,Evans2004}. Hence, the probability density
of the dissociation force for the double bond PSGL-1$-$P-selectin
complex is related to the single case by
\begin{eqnarray}
\label{equivalentbond} P_{\rm d}(f,f_0)=P\left(f/2,f_0/2\right)^2.
\end{eqnarray}
Fig.~\ref{figure3} presents the final result. We see that the
theoretical prediction agrees to the data very well. Hence we
conclude that the adiabatic approximation proposed at the
beginning is reasonable.
The previous works~\cite{Evans2004,Pereverzev} have claimed that
they could not fit the experimental data from the constant force
experiment using atomic force microscopy (AFM) and the force
jump-ramp experiment using BFP with the same parameters, {\it
e.g.}, see Tab. 1. The authors simply contributed it to the
different equipment and biological constructs though the
experiments studied the same complexes~\cite{Marshall,Evans2004}.
Our calculations however show that the mechanical parameters
defined by us have almost the same values. In addition, the
tendencies of our density functions for the first two panels of
the second array in Fig.~\ref{figure3} are closer to the data than
that predicted by the two-pathways
models~\cite{Evans2004,Pereverzev}.

The density functions and the force histograms in the experiments
reach the maximum and minimum at two distinct forces, which are
named $f_{\rm min}$ and $f_{\rm max}$ in the following,
respectively. This observation could be understood by setting the
derivative of Eq.~\ref{ruputureforcedistribution} with respect to
$f$ equal to zero,
\begin{eqnarray}
r\frac{d\lambda_0}{df}(f)=\lambda_0^2(f)\label{extremacondition},
\end{eqnarray}
We immediately see that the values of $f_{\rm min}$ and $f_{\rm
max}$ must be larger than the catch-slip transition force $f_c$
for the left term in Eq.~\ref{extremacondition} is negative as the
bond is catch. Indeed the experimental observations show that the
force values at the minimum histograms are around a certain values
even the loading rates change 10-fold. (see Figs.~2 and 4 in
Ref.~\cite{Evans2004}). The above equation could have no solutions
when the loading rate is smaller than a critical rate $r_c$, which
can be obtained by simultaneously solving
Eq.~\ref{extremacondition} and its first derivative. We estimate
$r_{\rm c}\approx 6$~pN/s using the current parameters, while the
force $f_{\rm max}=f_{\rm min}\approx 13$~pN (about 26~pN in the
double bond cases). If $r\le r_{\rm c}$, then the density function
is monotonous and decreasing function. Therefore, the most
probable force at the bond dissociation is zero. The most
interesting characteristics of Eq.~\ref{extremacondition} are the
dependence of the maximum and minimum forces on the loading rate.
In particular the latter is an important index in dynamic force
spectroscopy (DFS) theory since it corresponds to the most
possible dissociation force~\cite{Evans01}. When the loading rate
is sufficiently large, and correspondingly $f_{\rm max}$ is
larger, the approximation of Eq.~\ref{reactionratebendlandscape}
at large force Eq.~\ref{largeforcelimit} implies that
\begin{eqnarray}
f_{\rm max}&\approx& \frac{1}{\beta d_s}\ln\frac{\beta
rd^\ddag_s}{k_0^s\exp\left[\beta k_s^2/2k_x \right]}\propto \ln r.
\end{eqnarray}
The experimental measurement supported this prediction; see
Fig.~3A in Ref.~\cite{Evans2004}. On the other hand, due to that
$f_{\rm min}$ is very close to $f_c$ at the larger loading rate,
employing the Taylor's expanding approach we have
\begin{eqnarray}
\lambda_0(f)=\lambda_0(f_c)+
\frac{1}{2}\frac{d^2\lambda_0}{df^2}(f_c)(\delta f)^2+o[(\delta
f)^3],
\end{eqnarray}
where $\delta f=f-f_c$. Substituting it into
Eq.~\ref{extremacondition}, we get
\begin{eqnarray}
f_{\rm min}\approx
f_c+\lambda_0^2(f_c)\left/r\frac{d^2\lambda_0}{df^2}(f_c)\propto
r^{-1}.\right.
\end{eqnarray}
It means that $f_{\rm min}$ tends to $f_c$ very fast. Different
from $f_{\rm max}$, the loading rate dependence of $f_{\rm min}$
is an intrinsic property of the catch-slip bond; a unique
requirement is that the dissociation rate $\lambda_0(f)$ has a
minimum at the transition force $f_c$. Therefore $f_{\rm min}$s
observed in experiment performed by Evans {\it et
al.}~\cite{Evans2004} are almost the catch-slip transition force
observed in the constant force rupture experiment performed by
Marshall {\it et al.}~\cite{Marshall}. Indeed, the force values of
the minimum force histograms for the former are about~26 pN, while
the transition force for the latter ({\it dimeric}
PSGL-1$-$P-selectin) is also about 26~pN. We know that the
dissociation forces distribution of a simple slip bond only has a
maximum at a certain force value that depends on loading
rate~\cite{Izrailev}. Therefor the catch-slip bond can easily be
distinguished from the slip case by the presence of a minimum on
the density function of the dissociation forces at a nonvanished
force. Because the above analysis is independent of the initial
force $f_0$, in order to track the catch behaviors in the force
jump-ramp experiments, $f_0$ should be chosen to be smaller than
$f_c$.

\section{Conclusions and discussion}
Compared to the chemical kinetic schemes, our theory should be
more attractive on the following aspects. First of all, we suggest
that the counterintuitive catch-slip transition is a typical
example of the rate processes with dynamic disorder. Because this
concept has been broadly and deeply studied from theory and
experiment during the past two decades, extensive experience and
knowledge could be used for reference. For example, we suggest
that a new receptor-ligand forced dissociation experiment could be
performed over a large range of temperatures and solvent
viscosities. According to Eq.~\ref{origindiffusionequation}, if
the viscosity is so higher that $D\to 0$, we could predict
\begin{eqnarray}
p(x,t)\approx p(x,0)\exp[-tk_{\rm off}(x,f_{\parallel})].
\end{eqnarray}
We know that such a dissociation reaction is a typical example of
the rate processes with \emph{static} disorder~\cite{Zwanzig90}.
In addition that the survival probability of the bond converts
into multiple exponential decay at a single force from the single
exponential decay at the large $D$ limitation (see
Eq.~\ref{survivalprobability}), the mean lifetime is
\begin{eqnarray}
\langle\tau\rangle\approx\int p(x,0)k^{-1}_{\rm
off}(x,f_{\parallel}),
\end{eqnarray}
which means that the catch-slip bond changes into slip bond only.
Then our theory gives a intuitively obvious physical explanation
of catch bonds in an apparent expression
(Eq.~\ref{correctedBellform}): they could arise from a competition
of the two components of applied external force along the
dissociation reaction coordinate and the molecular conformational
coordinate; the former accelerates the dissociation by lowering
the height of the energy barrier, while the latter stabilizes the
complex by dragging the system to the higher barrier height.
Finally, the time-dependence of the forced dissociation rates
could be induced by either global conformational changes of the
complex or local conformational changes at the interface between
the receptor and ligand; no separated bound states and pathways
are needed in the current theory. Therefore it is possible that
one cannot find new stable complex structures through experiments
or detailed molecular dynamics (MD) simulations.

Even there are many advantages in the present theory. We cannot
definitely distinguish which theory or model is the most
reasonable and more close real situations with existing
experimental data. Moreover, except the coarse-grain physical
picture our theory does not reveal the detailed structural
information of the catch behavior of the ligand-receptor
complexes, while biologists might be more interested in it. We
could correspond the increasing height of the energy barrier with
respect to the conformational coordinate to the hook
structure~\cite{Isberg} or more affinity bound
states~\cite{Thomas}, however we believe that further
single-molecule experiments including micromanipulation
experiments and fluorescence spectroscopy, more crystal structure
data and detailed MD simulations from the atomic interactions are
essential to elucidate the real molecular mechanism of the catch
bonds.
\\
\\FL thanks Prof. Mian Long and Dr. Fei Ye for their helpful discussion
about the work.


\begin{thebibliography}{99}
\bibitem{McEver}
McEver, R. P. (2002). {\it Curr. Opin. Cell Biol.} {\bf 14},
581-586.

\bibitem{Konstantopoulos98}
Koonstantopoulos, K., Kurkreti, S. \& McIntire, L. V. (1998) {\it
Adv. Drug Deliv. Rev.} {\bf 33}, 141-164.


\bibitem{Marshall}
Marshall, B. T., Long, M., Piper, J. W., Yago, T., McIver, R. P.
\& Zhu, C. (2003) \nat{423}{190}{193}

\bibitem{Evans2004}
Evans, E., Leung, A., Heinrich, V., \& Zhu, C. (2004)
\pnas{101}{11281}{11286}.

\bibitem{Sarangapani}
Sarangapani, K. K., Yago, T., Klopocki, A. G., Lawrence, M. B.,
Fieger, C. B., Rosen, S. D., McEver, R. P. \& Zhu, C. (2003) {\it
J. Biol. Chem.} {\bf 279}, 2291-2298.

\bibitem{Yago}
Yago, T., Wu, J. H., Wey, C. D., Klopocki, A. G., Zhu, C., \&
McEver, R. P. (2004) {\it J. Cell. Biol.} {\bf 166}, 913-923.

\bibitem{Marshall05}
Marshall, B. T., Sarangapani, K. K., Lou, J. Z., McEver, R. P., \&
Zhu, C. (2005) \bio{88}{1458}{1466}.

\bibitem{Finger}
Finger, E. B., Puri, K. D., Alon, R. Lawrence, M. B., von Andrian,
U. H., \& Springer, T. A. (1996) \nat{279}{266}{269}.

\bibitem{Lawrence}
Lawrence, M. B., Kansas, G. S., Kunkel, E. J., \& Ley, K. (1997)
{\it J. Cell. Biol.} {\bf 136}, 717-727.


\bibitem{Bell}
Bell, G. I. (1978) \sci{200}{618}{627}.

\bibitem{Alon}
Alon, R., Hammer, D. A., \& Springer, T. A. (1995)
\nat{374}{539}{542}.

\bibitem{Chen}
Chen, S. \& Springer, T. A. (2003) \pnas{98}{950}{955}.

\bibitem{Dembo}
Dembo, M., Tourney, D. C., Saxman, K. \& Hammer, D. (1988) {\it
Proc. R. Soc. Lond.} B {\bf 234}, 55-83.

\bibitem{Dembo94}
Dembo, M. (1994) in {\it Lectures on Mathematics in the Life
Sciences: Some Mathematical Problems in Biology} eds. Goldstein,
B. \& Wofsy, C. (Am. Mathematical Soc., Providence, RI) Vol. {\bf
25}, pp. 1-27.

\bibitem{Evans97}
Evans, E. \& Ritchie, K. (1997) \bio{72}{1541}{1555}.

\bibitem{Barsegov}
Barsegov, V. \& Thirumalai, D. (2005) \pnas{102}{1835}{1840}.

\bibitem{Pereverzev}
Pereverzev, Y. V., Prezhdo, O. V., Forero, M., Sokurenko, E. V.,
\& Thomas, W. (2005) \bio{89}{1446}{1454}.

\bibitem{Somers}
Somers, W. S., Tang, J., Shaw, G. D., \& Camphausen, R. T. (2000)
\cell{103}{467}{479}.


\bibitem{Zwanzig90}
Zwanzig, R. (1990) {\it Acc. Chem. Res.} {\bf 23}, 148-152.

\bibitem{Agmon}
Agmon, N. \& Hopfield, J. J. (1983) \jcp{78}{6947}{6959}.


\bibitem{Kampen}
Van Kampen, N. G. J. Stat. Phys. (1977) {\bf 17}, 71-80.

\bibitem{Messiah}
Messiah, A. {\it Quantum mechanics} (North-Holland Pub. Co.
Amsterdam, 1962)


\bibitem{Hanley}
Hanley, W., McCarty, O. Jadhav, S., Tseng, Y., Wirtz, D. \&
Konstantopoulos, K. (2003) {\it J. Biol. Chem.} {\bf 278},
10556-10561.

\bibitem{Evans01}
Evans, E. A. (2001) {\it Annu. Rev. Biophys. Biomol. Struct.} {\bf
30}, 105-128.


\bibitem{Izrailev}
Izrailev, S., Stepaniants, s., Balsera, M., Oono, Y., \& Schulten,
K. (1997) \bio{72}{1568}{1581}.

\bibitem{Isberg}
Isberg, R. R. \& Barnes, P. (2002) \cell{110}{1}{4}.

\bibitem{Thomas}
Thomas, W. E., Trintchina, E., Forero, M., Vogel, V., \&
Sokurenko, E. V. (2002) \cell{109}{913}{923}.


\end{thebibliography}
 \end{document}